\documentclass[10pt]{article}

\usepackage{authblk}
\usepackage{natbib}
\usepackage{mathtools}
\usepackage{amsmath}
\usepackage{amsthm}
\usepackage{amssymb}
\usepackage{amsbsy}
\usepackage{amsfonts}
\usepackage{amscd}
\usepackage{mathrsfs}
\usepackage{bm}
\usepackage{breqn}
\usepackage{color, soul}
\usepackage{enumerate}
\usepackage{empheq}
\usepackage{rotating}
\usepackage{ftnxtra}
\usepackage{fnpos}
\usepackage[titletoc,toc,title]{appendix}
\usepackage{euscript}
\usepackage{graphicx}
\usepackage{epsfig}
\usepackage{epstopdf}
\DeclareGraphicsExtensions{.pdf,.png,.jpg,.eps}
\usepackage{pstool}
\usepackage{upgreek}
\usepackage{mathrsfs}

\usepackage{psfrag}

\numberwithin{equation}{section}

\theoremstyle{plain}	
\newtheorem{thm}{Theorem}[section]

\newtheorem{lem}[thm]{Lemma}
\newtheorem{prop}[thm]{Proposition}
\newtheorem*{prop*}{Proposition} 
\theoremstyle{definition}	
\newtheorem{definition}[thm]{Definition}
\newtheorem{remark}[thm]{Remark}
\newtheorem{example}[thm]{Example}

\usepackage{graphicx}
\usepackage{lipsum}

\usepackage{caption}
\usepackage{subcaption}

\usepackage{hyperref}
\hypersetup{colorlinks=true, linkcolor=blue}
\hypersetup{colorlinks=true,citecolor=blue}

\DeclareMathAlphabet{\mathpzc}{OT1}{pzc}{m}{it}

\usepackage{amsmath, amsthm, amssymb}

\usepackage{cleveref}

\DeclarePairedDelimiter\abs{\lvert}{\rvert}

\makeatletter
\newsavebox{\@brx}
\newcommand{\llangle}[1][]{\savebox{\@brx}{\(\m@th{#1\langle}\)}%
  \mathopen{\copy\@brx\mkern2mu\kern-0.9\wd\@brx\usebox{\@brx}}}
\newcommand{\rrangle}[1][]{\savebox{\@brx}{\(\m@th{#1\rangle}\)}%
  \mathclose{\copy\@brx\mkern2mu\kern-0.9\wd\@brx\usebox{\@brx}}}%
\let\oldabs\abs
\def\abs{\@ifstar{\oldabs}{\oldabs*}}
\makeatother

\usepackage{accents}

    {\end{bmatrix}}%

\usepackage{enumitem}

\usepackage[utf8]{inputenc}
\usepackage{dutchcal}
\usepackage{multirow}

\usepackage{xcolor}

\usepackage{setspace}

\begin{document}

\title{\textbf{Controllable Deformations in Compressible\\ Isotropic Implicit Elasticity}}

\author[1,2]{Arash Yavari\thanks{Corresponding author, e-mail: arash.yavari@ce.gatech.edu}}
\author[3]{Alain Goriely}
\affil[1]{\small \textit{School of Civil and Environmental Engineering, Georgia Institute of Technology, Atlanta, GA 30332, USA}}
\affil[2]{\small \textit{The George W. Woodruff School of Mechanical Engineering, Georgia Institute of Technology, Atlanta, GA 30332, USA}}
\affil[3]{\small \textit{Mathematical Institute, University of Oxford, Oxford, OX2 6GG, UK}}

\maketitle

\begin{abstract}
For a given material, \emph{controllable deformations} are those deformations that can be maintained in the absence of body forces and by applying only boundary tractions.
For a given class of materials, \emph{universal deformations} are those deformations that are controllable for any material within the class. In this paper, we characterize the universal deformations in compressible isotropic implicit elasticity defined by solids whose constitutive equations, in terms of the Cauchy stress $\boldsymbol{\sigma}$ and the left Cauchy-Green strain $\mathbf{b}$, have the implicit form $\boldsymbol{\mathsf{f}}(\boldsymbol{\sigma},\mathbf{b})=\mathbf{0}$. 
We prove that universal deformations are homogeneous. However, an important observation is that, unlike Cauchy (and Green) elasticity, not every homogeneous deformation is permissible for a given implicit-elastic solid. In other words, the set of universal deformations is material-dependent, yet it remains a subset of homogeneous deformations.
\end{abstract}

\begin{description}
\item[Keywords:] Universal deformation, implicit constitutive equations, Cauchy elasticity, hyperelasticity, Green elasticity, isotropic solids.
\end{description}


\section{Introduction}

For a specific class of materials, universal deformations are those deformations that can be maintained in the absence of body forces, solely by applying boundary tractions, for all members of the material class.
For Cauchy elastic (and particularly, hyperelastic) solids, universal deformations are independent of the specific material within the class. However, the boundary tractions required to maintain a universal deformation explicitly depend on the particular material.
Universal deformations have played a pivotal role in nonlinear elasticity and anelasticity. The following are some examples:
\begin{itemize}[topsep=2pt,noitemsep, leftmargin=10pt]
\item They have played a crucial organizational role in the semi-inverse solutions in nonlinear elasticity \citep{Knowles1979,Polignone1991,DePascalis2009,Tadmor2012,Goriely2017}, and more recently in anelasticity \citep{KumarYavari2023} and viscoelasticity \citep{SadikYavari2024}. 
\item They provide guidance for designing experiments aimed at determining the constitutive relations of a specific material \citep{Rivlin1951,DoyleEricksen1956,Saccomandi2001}.
\item All the existing exact solutions for defects in nonlinear solids are associated with universal deformations \citep{Wesolowski1968,Gairola1979,Zubov1997,YavariGoriely2012a,YavariGoriely2012b,YavariGoriely2013a,YavariGoriely2014,Golgoon2018}.
\item Universal deformations have been crucial in finding exact solutions for distributed finite eigenstrains in nonlinear solids, as well as in solving the nonlinear analogues of Eshelby's inclusion problem \citep{YavariGoriely2013b,YavariGoriely2015,Golgoon2017,Yavari2021Eshelby}.
\item Universal deformations are exact solutions that have served as benchmark problems in computational mechanics \citep{Dragoni1996,Saccomandi2001,Chi2015,Shojaei2018}.
\item Universal deformations have been utilized in deriving effective properties for nonlinear composites \citep{Hashin1985,Lopez2012,Golgoon2021}.
\end{itemize}

The systematic investigation of universal deformations for homogeneous compressible and incompressible isotropic hyperelastic solids began in the 1950s with the seminal works of Jerry Ericksen \citep{Ericksen1954,Ericksen1955} whose work was inspired by the earlier contributions of Ronald Rivlin \citep{Rivlin1948, Rivlin1949a, Rivlin1949b}. \citet{Ericksen1955} demonstrated that in homogeneous compressible isotropic solids, universal deformations are homogeneous.
Characterizing universal deformations in the presence of internal constraints such as incompressibility poses a particularly challenging problem \citep{Saccomandi2001}. \citet{Ericksen1954} identified four families of universal deformations for incompressible isotropic hyperelastic solids. He conjectured that a deformation with constant principal invariants is homogeneous, but this conjecture turned out to be incorrect \citep{Fosdick1966}.
Later on, a fifth family of universal deformations was found \citep{SinghPipkin1965,KlingbeilShield1966}.
The fifth family of universal deformations are inhomogeneous, yet these deformations have constant principal invariants.  To this date, the problem of whether there exist additional inhomogeneous constant-principal invariant universal deformations remains open (Ericksen's problem).

The investigation into universal deformations was expanded to inhomogeneous anisotropic solids in \citep{YavariGoriely2021,Yavari2021,YavariGoriely2023Universal}.
Before our studies, there had been limited research on universal deformations in anisotropic solids \citep{Ericksen1954Anisotropic}.
The comprehensive analyses presented in \citep{YavariGoriely2021,Yavari2021,YavariGoriely2023Universal} encompassed both compressible and incompressible isotropic, transversely isotropic, orthotropic, and monoclinic solids. It was demonstrated that for these three classes of compressible anisotropic solids, universal deformations are homogeneous, and the preferred material directions are uniform. Moreover, for isotropic solids and each of the three classes of anisotropic solids, the corresponding universal inhomogeneities, which represent inhomogeneities in the energy function compatible with the universality constraints, were characterized.
For inhomogeneous incompressible isotropic and the three classes of inhomogeneous incompressible anisotropic solids, the corresponding universal inhomogeneities for each of the above six known families of universal deformations were determined.

In linear elasticity, universal displacements are counterparts to universal deformations \citep{Truesdell1966,Gurtin1972,Carroll1973,Yavari2020}. \citet{Yavari2020} showed the explicit dependence of universal displacements on the symmetry class of the material. In particular, the larger the symmetry group, the larger the corresponding set of universal displacements. 
The study of universal displacements has also been expanded to include inhomogeneous solids
\citep{YavariGoriely2022}, linear anelasticity \citep{Yavari2022Anelastic-Universality}, and compressible anisotropic linear elastic solids reinforced by a single family of inextensible fibers \citep{Yavari2023}.

Recently, there have been extensions of Ericksen's analysis to include anelasticity.
\citet{YavariGoriely2016} proved that in compressible anelasticity, universal deformations are covariantly homogeneous, and for simply-connected bodies universal eigenstrains are impotent (zero-stress). 
A partial characterization of universal deformations and eigenstrains in incompressible anelasticity was given in \citep{Goodbrake2020}.
In particular, it was observed that the six known families of universal deformations for incompressible, isotropic elastic solids are invariant under specific Lie subgroups of the special Euclidean group.
There are also recent studies on universal deformations and eigenstrains in accreting bodies \citep{YavariPradhan2022,YavariAccretion2023,PradhanYavari2023} and investigations into universal deformations in liquid crystal elastomers \citep{LeeBhattacharya2023,MihaiGoriely2023}.

In a recent study, \citet{Yavari2024} extended the analysis of universal deformations and inhomogeneities to inhomogeneous compressible and incompressible isotropic Cauchy elasticity.
In Cauchy elasticity, which includes hyperelasticity (Green elasticity) as a special case, an energy function does not generally exist.
It was demonstrated that somewhat unexpectedly the sets of universal deformations and inhomogeneities of Cauchy elasticity are identical to those of Green elasticity in both compressible and incompressible cases.

Here, we study universal deformations in compressible isotropic solids with implicit constitutive equations of the form $\boldsymbol{\mathsf{f}}(\boldsymbol{\sigma},\mathbf{b})=\mathbf{0}$, where $\boldsymbol{\sigma}$ is the Cauchy stress and $\mathbf{b}$ is the left Cauchy-Green strain. This class of materials was introduced by \citet{Morgan1966} who used a result due to \citet{RivlinEricksen1955} to simplify such implicit constitutive equations for isotropic solids. 
This work remained largely unnoticed until Rajagopal and his collaborators started a major research program on the mechanics of elastic solids with implicit constitutive equations in the past two decades \citep{Rajagopal2003,Rajagopal2007,Bustamante2009,Bustamante2011}.

This paper is organized as follows. 
In \S\ref{Imp-Elasticity}, implicit elasticity is briefly reviewed. Universal deformations of compressible isotropic implicit-elastic solids are characterized in \S\ref{C-Elasticity-UC}. 
Conclusions are given in \S\ref{Sec:Conclusions}.

\section{Implicit elasticity} \label{Imp-Elasticity}

Consider a body that in its undeformed configuration is identified with an embedded submanifold $\mathcal{B}$ of the Euclidean ambient space $\mathcal{S}$. The flat metric of the Euclidean ambient space is denoted by $\mathbf{g}$ and the induced metric on the body in its reference configuration is denoted by $\mathbf{G}=\mathbf{g}|_{\mathcal{B}}$. A \textit{deformation} is a map from $\mathcal{B}$ to the ambient space, i.e., $\varphi:\mathcal{B}\to\mathcal{C}\subset\mathcal{S}$, where $\mathcal{C}=\varphi(\mathcal{B})$ is the current configuration. The tangent map of $\varphi$ is the so-called \textit{deformation gradient} $\mathbf{F}=T\varphi$ (a metric-independent map), which at each material point $X \in \mathcal{B}$ is a linear map $\mathbf{F}(X):T_{X}\mathcal{B}\rightarrow T_{\varphi(X)}\mathcal{C}$. 
With respect to the coordinate charts $\{X^A\}$ and $\{x^a\}$ for $\mathcal{B}$ and $\mathcal{C}$, respectively, the deformation gradient has components $F^a{}_A=\frac{\partial \varphi^a}{\partial X^A}$.
The transpose of deformation gradient $\mathbf{F}^{\textsf{T}}$ has components $(F^{\textsf{T}})^A{}_{a}=g_{ab}\,F^b{}_{B}\,G^{AB}$.
The \textit{right Cauchy-Green strain tensor} is defined as $\mathbf{C}=\mathbf{F}^{\textsf{T}}\mathbf{F}$ with components $C^A{}_{B}=(F^{\textsf{T}})^A{}_{a}F^a{}_{B}$.
Thus, $C_{AB}=(g_{ab}\circ \varphi)\,F^a{}_{A}\,F^b{}_{B}$, which means that the right Cauchy-Green strain is  the pull-back of the spatial metric to the reference configuration, i.e., $\mathbf{C}^\flat=\varphi^*\mathbf{g}$, where $\flat$ is the flat operator induced by the metric $\mathbf{G}$ (which lowers indices).
The left Cauchy-Green strain is defined as $\mathbf{B}^{\sharp}=\varphi^*(\mathbf{g}^{\sharp})$, which has components $B^{AB}=F^{-A}{}_a\,F^{-B}{}_b\,g^{ab}$. 
The spatial analogue of $\mathbf{C}^\flat$ is defined as $\mathbf{c}^\flat=\varphi_*\mathbf{G}$, which has components $c_{ab}=F^{-A}{}_a\,F^{-B}{}_b\,G_{AB}$.
Similarly, the spatial analogue of $\mathbf{B}^{\sharp}$ is $\mathbf{b}^{\sharp}=\varphi_*(\mathbf{G}^{\sharp})$, which has components $b^{ab}=F^a{}_{A}\,F^b{}_{B}\,G^{AB}$.
Recall that $\mathbf{b}=\mathbf{c}^{-1}$.
The two tensors $\mathbf{C}$ and $\mathbf{b}$ share the same principal invariants $I_1$, $I_2$, and $I_3$, which are defined as \citep{Ogden1984,MarsdenHughes1994} $I_1 =\operatorname{tr}\mathbf{b}=b^{ab}\,g_{ab}$, $I_2=\frac{1}{2}\left(I_1^2-\operatorname{tr}\mathbf{b}^2\right)=\frac{1}{2}\left(I_1^2-b^{ab}b^{cd}\,g_{ac}\,g_{bd}\right)$, and $I_3=\det \mathbf{b}$.

Implicit elasticity is defined by  elastic materials that have the following implicit constitutive equation \citep{Morgan1966,Rajagopal2003,Rajagopal2007}\footnote{In his study of \emph{controllable states of stress}, \citet{Carroll1973} considered a special subset of \eqref{ImplicitStressStrain} constitutive equations in the form 
\begin{equation} \label{invertible-stress}
	\mathbf{b}^\sharp=\xi_0\,\mathbf{g}^\sharp+\xi_1 \boldsymbol{\sigma}+\xi_2 \boldsymbol{\sigma}^2\,,
\end{equation}
that he called an \emph{invertible stress relation}. 
The response function $\xi_0$, $\xi_1$, and $\xi_2$ are functions of the principal invariants of the Cauchy stress. According to \citet{Carroll1973}, a state of stress is controllable if it satisfies the equilibrium equations in the absence of body forces, and its corresponding strain is compatible for any response functions $\xi_0$, $\xi_1$, and $\xi_2$. He proved that for isotropic elastic solids with constitutive equations of the form \eqref{invertible-stress} controllable states of stress are homogeneous. He, however, noted that for a given elastic material with the constitutive equation \eqref{invertible-stress}, not all states of homogeneous stress are admissible.}
\begin{equation} \label{ImplicitStressStrain}
	\boldsymbol{\mathsf{f}}(\boldsymbol{\sigma},\mathbf{b})=\mathbf{0}\,. 
\end{equation}
Cauchy elastic and Green elastic (hyperelastic) solids are special cases of this class of elastic materials.
Here, we confine ourselves to isotropic solids. In this case, the implicit constitutive equation can be rewritten as \citep{RivlinEricksen1955,Morgan1966}\footnote{In \citep{RivlinEricksen1955,Morgan1966}, this is written as
\begin{equation}
\begin{aligned}
	&\beta_0\,\mathbf{g}^\sharp
	+\beta_1 \boldsymbol{\sigma}+\beta_2 \boldsymbol{\sigma}^2
	+\beta_3\,\mathbf{b}^\sharp+\beta_4\,\mathbf{b}^{2\sharp} 
	+\beta_5 \left(\boldsymbol{\sigma}\mathbf{b}^\sharp
	+\mathbf{b}^\sharp\boldsymbol{\sigma}\right)
	+\beta_6 \left(\boldsymbol{\sigma}^2\mathbf{b}^\sharp
	+\mathbf{b}^\sharp\boldsymbol{\sigma}^2\right)\\
	&
	+\beta_7 \left(\boldsymbol{\sigma}\mathbf{b}^{2\sharp}
	+\mathbf{b}^{2\sharp}\boldsymbol{\sigma}\right)
	+\beta_8 \left(\boldsymbol{\sigma}^2\mathbf{b}^{2\sharp}
	+\mathbf{b}^{2\sharp}\boldsymbol{\sigma}^2\right)
	=\mathbf{0}\,.
\end{aligned}
\end{equation}
However, from the Cayley-Hamilton theorem we know that $\mathbf{b}^{2\sharp}$ is fuctionally dependent on $\mathbf{b}^{\sharp}$ and $\mathbf{b}^{-\sharp}=\mathbf{c}^{\sharp}$.
}
\begin{equation} \label{implicit-sigma-b}
\begin{aligned}
	&\beta_0\,\mathbf{g}^\sharp
	+\beta_1 \boldsymbol{\sigma}+\beta_2 \boldsymbol{\sigma}^2
	+\beta_3\,\mathbf{b}^\sharp+\beta_4\,\mathbf{c}^{\sharp} 
	+\beta_5 \left(\boldsymbol{\sigma}\mathbf{b}^\sharp
	+\mathbf{b}^\sharp\boldsymbol{\sigma}\right)
	+\beta_6 \left(\boldsymbol{\sigma}^2\mathbf{b}^\sharp
	+\mathbf{b}^\sharp\boldsymbol{\sigma}^2\right) \\
	&
	+\beta_7 \left(\boldsymbol{\sigma}\mathbf{c}^{\sharp}
	+\mathbf{c}^{\sharp}\boldsymbol{\sigma}\right)+\beta_8 \left(\boldsymbol{\sigma}^2\mathbf{c}^{\sharp}
	+\mathbf{c}^{\sharp}\boldsymbol{\sigma}^2\right)
	=\mathbf{0}\,,
\end{aligned}
\end{equation}
where $\beta_i,~i=0,\hdots,8$  are functions of the following ten invariants \citep{RivlinEricksen1955,Morgan1966,Rajagopal2003,Rajagopal2007}:
\begin{equation}
	\operatorname{tr}\boldsymbol{\sigma}\,,~~
	\operatorname{tr}\boldsymbol{\sigma}^2\,,~~
	\operatorname{tr}\boldsymbol{\sigma}^3\,,~~
	\operatorname{tr}\mathbf{b}^{\sharp} \,,~~
	\operatorname{tr}\mathbf{b}^{2\sharp} \,,~~
	\operatorname{tr}\mathbf{b}^{3\sharp} \,,~~
	\operatorname{tr}\left(\boldsymbol{\sigma}\mathbf{b}^{\sharp}\right)\,,~~
	\operatorname{tr}\left(\boldsymbol{\sigma}\mathbf{b}^{2\sharp}\right)\,,~~
	\operatorname{tr}\left(\boldsymbol{\sigma}^2\mathbf{b}^{\sharp}\right)\,,~~
	\operatorname{tr}\left(\boldsymbol{\sigma}^2\mathbf{b}^{2\sharp}\right)
	\,.
\end{equation}
Equivalently, we can  use the following invariants\footnote{Clearly, one can use the pair $I_4$ and $I_5$ instead of $I_4$ and $\operatorname{tr}\mathbf{b}^{2\sharp}$. Using the Calyley-Hamilton theorem, one can  show that $\operatorname{tr}\mathbf{b}^{3\sharp}=I_4^3-3I_4I_5+3I_6$ \citep{Spencer1971}.}
\begin{equation} \label{I-Invariants}
\begin{aligned}
	& I_1=\operatorname{tr}\boldsymbol{\sigma}\,,\quad
	I_2=\operatorname{tr}\boldsymbol{\sigma}^2\,,\quad
	I_3=\operatorname{tr}\boldsymbol{\sigma}^3\,,\quad
	I_4=\operatorname{tr}\mathbf{b}^{\sharp} \,,\quad
	I_5=\frac{1}{2}\left[ I_4^2- \operatorname{tr}\mathbf{b}^{2\sharp}\right]\,,\quad
	I_6=\det\mathbf{b}^{\sharp} \,,\quad \\
	& I_7=\operatorname{tr}\left(\boldsymbol{\sigma}\mathbf{b}^{\sharp}\right)\,,\quad
	I_8=\operatorname{tr}\left(\boldsymbol{\sigma}\mathbf{b}^{2\sharp}\right)\,,\quad
	I_9=\operatorname{tr}\left(\boldsymbol{\sigma}^2\mathbf{b}^{\sharp}\right)\,,\quad
	I_{10}=\operatorname{tr}\left(\boldsymbol{\sigma}^2\mathbf{b}^{2\sharp}\right)
	\,.
\end{aligned}
\end{equation}
Hence, the implicit constitutive equations are rewritten as\footnote{Assuming that the initial configuration is stress free, $\alpha_0+\alpha_3+\alpha_4=0$ when $(I_1,\hdots,I_{10})=(0,0,0,3,3,1,0,0,0,0)$.}
\begin{equation} \label{Implicit-Constitutive-Equation-Isotropic}
\begin{aligned}
	\boldsymbol{\mathsf{f}}(\boldsymbol{\sigma},\mathbf{b})
	=&\alpha_0\,\mathbf{g}^\sharp
	+\alpha_1 \boldsymbol{\sigma}+\alpha_2 \boldsymbol{\sigma}^2
	+\alpha_3\,\mathbf{b}^\sharp+\alpha_4\,\mathbf{c}^{\sharp} 
	+\alpha_5 \left(\boldsymbol{\sigma}\mathbf{b}^\sharp
	+\mathbf{b}^\sharp\boldsymbol{\sigma}\right)
	+\alpha_6 \left(\boldsymbol{\sigma}^2\mathbf{b}^\sharp
	+\mathbf{b}^\sharp\boldsymbol{\sigma}^2\right)\\
	&
	+\alpha_7 \left(\boldsymbol{\sigma}\mathbf{c}^{\sharp}
	+\mathbf{c}^{\sharp}\boldsymbol{\sigma}\right)
	+\alpha_8 \left(\boldsymbol{\sigma}^2\mathbf{c}^{\sharp}
	+\mathbf{c}^{\sharp}\boldsymbol{\sigma}^2\right)
	=\mathbf{0}\,,
\end{aligned}
\end{equation}
where $\alpha_i=\alpha_i(I_1,\hdots,I_{10})$, $i=0,\hdots,8$.
Note that with respect to a coordinate chart $\{x^a\}$, \eqref{Implicit-Constitutive-Equation-Isotropic} is written as ($a,b=1,2,3$)
\begin{equation}\label{Implicit-Constitutive-Equation-Isotropic-Comp}
\begin{aligned}
	&\alpha_0\,g^{ab}
	+\alpha_1 \sigma^{ab}+\alpha_2\, \sigma^{2ab}
	+\alpha_3\,b^{ab}+\alpha_4\,c^{ab} 
	+\alpha_5 \left(\sigma^{am}\,b_m^b	+b^a_m\,\sigma^{mb}\right)
	+\alpha_6 \left(\sigma^{2am}\,b_m^b	+b^a_m\,\sigma^{2mb}\right)\\
	&
	+\alpha_7 \left(\sigma^{am}\,c_m^{b}+c^{a}_m\,\sigma^{mb}\right)
	+\alpha_8 \left(\sigma^{2am}\,c_m^{b}+c^{a}_m\,\sigma^{2mb}\right)
	=0\,,
\end{aligned}
\end{equation}
where $\sigma^{2ab}=\sigma^{am}\,g_{mn}\,\sigma^{nb}$.

\begin{definition}
The \emph{implicit-elasticity stress-strain space} $\mathbb{S}$ is a twelve-dimensional submanifold of $\mathbb{R}^{12}$
defined as
\begin{equation}
	\mathbb{S}=\left\{s\in\mathbb{R}^{12}: s=(b^{ab},\sigma^{cd})\,,~a\leq b=1,2,3\,,~c\leq d=1,2,3\,,
	~ \mathbf{b}~\text{positive-definite~and~compatible}  \right\}.
\end{equation}
The \emph{implicit-elasticity stress-strain manifold} $\mathfrak{S}$ of an implicit-elastic material is the six-dimensional submanifold of $\mathbb{S}$ defined by the six equations $\boldsymbol{\mathsf{f}}(\boldsymbol{\sigma},\mathbf{b})=\mathbf{0}$.
\end{definition}

\begin{definition}
In implicit elasticity, a homogeneous deformation (that has a constant $\mathbf{b}$) is \textit{admissible} if $(\mathbf{b},\boldsymbol{\sigma})\in\mathfrak{S}$, i.e., if there exists $\boldsymbol{\sigma}$ such that $\boldsymbol{\mathsf{f}}(\boldsymbol{\sigma},\mathbf{b})=\mathbf{0}$.
\end{definition}

\begin{example}\label{Ex:quadratic}
In \eqref{Implicit-Constitutive-Equation-Isotropic} suppose $\alpha_5=\alpha_6=\alpha_7=\alpha_8=0$, i.e.,
\begin{equation}
	\alpha_2 \boldsymbol{\sigma}^2+\alpha_1 \boldsymbol{\sigma}
	=-\alpha_0\,\mathbf{g}^\sharp-\alpha_3\,\mathbf{b}^\sharp-\alpha_4\,\mathbf{c}^{\sharp} \,.
\end{equation}
We denote the eigenvalues of $\boldsymbol{\sigma}$ and $\mathbf{b}^\sharp$ by $\sigma_i$ and $\lambda_i^2$, $i=1,2,3$, respectively.
For this class of materials, $\boldsymbol{\sigma}$ and $\mathbf{b}^\sharp$ have the same eigenvectors and can be diagonalized simultaneously. Therefore, with respect to the eigenbasis one has
\begin{equation}
	\alpha_2 \sigma_i^2+\alpha_1 \sigma_i+(\alpha_0+\alpha_3\,\lambda_i^2+\alpha_4\,\lambda_i^{-2})=0
	\,,\qquad i=1,2,3 \,.
\end{equation}
These quadratic equations (in $\sigma_i$) have real solutions if and only if $\alpha_1^2-4\alpha_2(\alpha_0+\alpha_3\,\lambda_i^2+\alpha_4\,\lambda_i^{-2})\geq 0$, $i=1,2,3$. Thus, $\mathbf{b}^\sharp$ is inadmissible if for at least one value of $i$ one has
\begin{equation}
	\alpha_1^2-4\alpha_2\left(\alpha_0+\alpha_3\,\lambda_i^2+\alpha_4\,\lambda_i^{-2}\right) < 0 \,.
\end{equation}
We observe that not every homogeneous deformation is admissible.
\end{example}

\begin{example}
Consider the particular case of  \eqref{Implicit-Constitutive-Equation-Isotropic} for which  $\alpha_1=\alpha_2=\alpha_6=\alpha_7=\alpha_8=0$ and $\alpha_5\neq 0$ , i.e.,
\begin{equation} \label{Sylvester-equation}
	\boldsymbol{\sigma}\mathbf{b}^\sharp
	+\mathbf{b}^\sharp\boldsymbol{\sigma}=\eta_0\,\mathbf{g}^\sharp
	+\eta_1\,\mathbf{b}^\sharp+\eta_2\,\mathbf{c}^{\sharp} \,.
\end{equation}
For a homogeneous deformation and constant $\eta_0, \eta_1, \eta_2$ (these are constant when stress is homogeneous as well), \eqref{Sylvester-equation} is a Sylvester equation \citep{Sylvester1884,Bhatia1997,Gantmakher2000}. There is a unique solution for stress if and only if $\mathbf{b}^\sharp$ and $-\mathbf{b}^\sharp$ do not share any eigenvalues. This is the case because all the eigenvalues of $\mathbf{b}^\sharp$ are positive and consequently those of $-\mathbf{b}^\sharp$ are all negative. Therefore, any homogeneous $\mathbf{b}^\sharp$ has a unique corresponding homogeneous stress.
\end{example}

\section{Universal Deformations in Compressible Isotropic Implicit Elasticy} \label{C-Elasticity-UC}

\paragraph{Universal deformations for a special class of Cauchy-elastic solids.}
Before considering the general case of implicit elasticity, we start by characterizing the universal deformations of isotropic Cauchy-elastic solids. This class of material is defined by an explicit constitutive equations of the form:
\begin{equation}
	\boldsymbol{\sigma}=\beta_0\,\mathbf{g}^\sharp
	+\beta_1\,\mathbf{b}^\sharp+\beta_2\,\mathbf{c}^{\sharp} \,,
\end{equation}
where $\beta_i=\beta_i(I_1,I_2,I_3)$, $i=1,2,3$.
Equilibrium equations in the absence of body forces are written as (note that metric is covariantly constant, and hence, $\operatorname{div}\mathbf{g}^\sharp=\mathbf{0}$)
\begin{equation}
	\mathbf{0}=\operatorname{div}\boldsymbol{\sigma}
	=\operatorname{div}\left[\beta_0\,\mathbf{g}^\sharp
	+\beta_1\,\mathbf{b}^\sharp+\beta_2\,\mathbf{c}^{\sharp}\right]
	=\beta_1\operatorname{div}\mathbf{b}^\sharp
	+\beta_2 \operatorname{div}\mathbf{c}^{\sharp}
	+\nabla\beta_0
	+\mathbf{b}^\sharp\cdot \nabla\beta_1
	+\mathbf{c}^{\sharp}\cdot\nabla\beta_2 \,.
\end{equation}
Then, it is noted that
\begin{equation}
	\nabla \beta_i=\sum_{j=1}^3\beta_{ij}\nabla I_j\,,\qquad
	\beta_{ij}=\frac{\partial \beta_{i}}{\partial I_j}\,,\quad i=1,2,3\,.
\end{equation}
Thus, we have
\begin{equation} \label{Universality-Cauchy}
	\beta_1\operatorname{div}\mathbf{b}^\sharp
	+\beta_2 \operatorname{div}\mathbf{c}^{\sharp}
	+\sum_{j=1}^3\beta_{0j}\nabla I_j
	+\sum_{j=1}^3\beta_{1j} \mathbf{b}^\sharp\cdot\nabla I_j
	+\sum_{j=1}^3\beta_{2j} \mathbf{c}^\sharp\cdot\nabla I_j
	=\mathbf{0}\,.
\end{equation}
This identity must hold for arbitrary response functions $\beta_0$, $\beta_1$, and $\beta_2$. Knowing that any derivative of a response function is functionally independent from the response functions and all the other derivatives, one concludes that in \eqref{Universality-Cauchy} the coefficient of each response function and its partial derivatives must be identically zero. This immediately implies that \citep{Yavari2024}
\begin{equation} \label{Universality-Constraints-Cauchy}
	\operatorname{div}\mathbf{b}^\sharp=\operatorname{div}\mathbf{c}^{\sharp}=\mathbf{0}\,,
	\qquad \text{and}\quad \operatorname{grad}I_j=\mathbf{0}\,,~i=1,2,3\,.
\end{equation}
These constraints are identical to the universality constraints of isotropic hyperelasticity and imply that the universal deformations are homogeneous \citep{Ericksen1955}.
It should be emphasized that universal deformations are independent of the response functions $\beta_0$, $\beta_1$, and $\beta_2$. However, for a homogeneous body with a given triplet of response functions $(\beta_0,\beta_1,\beta_2)$, stress explicitly depends on the response functions, but is nevertheless  uniform.

\paragraph{Equilibrium stress with constant principal invariants is homogeneous.}

To extend this analysis to the more general case of isotropic implicit elasticity given by  constitutive equations \eqref{Implicit-Constitutive-Equation-Isotropic} we will need  the following general result. 

\begin{lem}
If $\boldsymbol{\sigma}$ is an equilibrium stress, i.e. $\operatorname{div}\boldsymbol{\sigma}=\mathbf{0}$, with constant principal invariants, then it is homogeneous.
\end{lem}
\begin{proof}
The Cauchy stress can be written in the following spectral representation \citep{Ogden1984}:
\begin{equation}
	\boldsymbol{\sigma}= 
	\sigma_1 \,\boldsymbol{\mathsf{a}}\otimes\boldsymbol{\mathsf{a}}	
	+\sigma_2 \,\boldsymbol{\mathsf{b}}\otimes\boldsymbol{\mathsf{b}}
	+\sigma_3 \,\boldsymbol{\mathsf{c}}\otimes\boldsymbol{\mathsf{c}}
	\,,
\end{equation}
where $\sigma_1, \sigma_2, \sigma_3$ and $\boldsymbol{\mathsf{a}}, \boldsymbol{\mathsf{b}}, \boldsymbol{\mathsf{c}}$ are the principal stresses and their corresponding principal directions. Note that $\boldsymbol{\mathsf{a}}\otimes\boldsymbol{\mathsf{a}}+ \boldsymbol{\mathsf{b}}\otimes\boldsymbol{\mathsf{b}}+ \boldsymbol{\mathsf{c}}\otimes\boldsymbol{\mathsf{c}}=\mathbf{g}^\sharp$.
Thus, the Cauchy stress can be written
\begin{equation} \label{Stress-Specrtal}
	\boldsymbol{\sigma}= 
	(\sigma_1-\sigma_3) \,\boldsymbol{\mathsf{a}}\otimes\boldsymbol{\mathsf{a}}	
	+(\sigma_2-\sigma_3) \,\boldsymbol{\mathsf{b}}\otimes\boldsymbol{\mathsf{b}}
	+\sigma_3 \,\mathbf{g}^\sharp	\,.
\end{equation}
If it has constant invariants then it has constant eigenvalues. Moreover, being an equilibrium stress,  it is divergence-free. Therefore, there are three possibilities for the eigenvalues: \textit{(i) }$\sigma_1=\sigma_2=\sigma_3$, \textit{(ii)} $\sigma_2=\sigma_3\neq \sigma_1$, and \textit{(iii)} the three eigenvalues are distinct.
Case \textit{(i)}: When the three eigenvalues of $\boldsymbol{\sigma}$ are equal, we have $\boldsymbol{\sigma}=\sigma_1 \,\mathbf{g}^\sharp$, which is a covariantly constant tensor, i.e., a homogeneous tensor in the Euclidean space. \\
Case \textit{(ii)}: when $\sigma_2=\sigma_3\neq \sigma_1$, the stress has the following representation
\begin{equation}
	\boldsymbol{\sigma}= 
	(\sigma_1-\sigma_3) \,\boldsymbol{\mathsf{a}}\otimes\boldsymbol{\mathsf{a}}	
	+\sigma_3 \,\mathbf{g}^\sharp	\,.
\end{equation}
Thus, $\operatorname{div}\boldsymbol{\sigma}=(\sigma_1-\sigma_3)\operatorname{div}(\boldsymbol{\mathsf{a}}\otimes\boldsymbol{\mathsf{a}})=\mathbf{0}$, and hence $\operatorname{div}(\boldsymbol{\mathsf{a}}\otimes\boldsymbol{\mathsf{a}})=\mathbf{0}$. Note that
\begin{equation}
	\operatorname{div}(\boldsymbol{\mathsf{a}}\otimes\boldsymbol{\mathsf{a}})
	=\nabla_{\boldsymbol{\mathsf{a}}}\,\boldsymbol{\mathsf{a}}
	+(\operatorname{div}\boldsymbol{\mathsf{a}})\,\boldsymbol{\mathsf{a}}\,,\quad \text{or in components } 
	\quad (\mathsf{a}^m\mathsf{a}^n)_{|n}=\mathsf{a}^m{}_{|n}\mathsf{a}^n
	+\mathsf{a}^n{}_{|n}\,\mathsf{a}^m
		\,.
\end{equation}
Therefore, equilibrium dictates that
\begin{equation}  \label{Equilibrium-a}
	\nabla_{\boldsymbol{\mathsf{a}}}\,\boldsymbol{\mathsf{a}}
	+(\operatorname{div}\boldsymbol{\mathsf{a}})\,\boldsymbol{\mathsf{a}}=\mathbf{0}\,,
	\quad \text{or in components } \quad
	\mathsf{a}^m{}_{|n}\mathsf{a}^n+\mathsf{a}^n{}_{|n}\,\mathsf{a}^m=0		\,.
\end{equation}
Knowing that $\mathsf{a}^m\mathsf{a}_m=1$, one has $0=\left(\mathsf{a}^m\mathsf{a}_m\right)_{|b}=\mathsf{a}^m{}_{|b}\mathsf{a}_m+\mathsf{a}^m\mathsf{a}_m{}_{|b}=\mathsf{a}^m{}_{|b}\mathsf{a}_m+\mathsf{a}_m\mathsf{a}^m{}_{|b}=2\mathsf{a}^m{}_{|b}\mathsf{a}_m$. Thus, for any unit vector $\boldsymbol{\mathsf{a}}$
\begin{equation} \label{Identity-Unit-a}
	\mathsf{a}_m\mathsf{a}^m{}_{|b}=0\,.
\end{equation}
Taking the dot product of the vector equation \eqref{Equilibrium-a} with  $\boldsymbol{\mathsf{a}}$ and using \eqref{Identity-Unit-a}, one obtains
\begin{equation}  
	0=\mathsf{a}_m \mathsf{a}^m{}_{|n}\mathsf{a}^n+\mathsf{a}^n{}_{|n}\,\mathsf{a}^m \mathsf{a}_m
	=\mathsf{a}^n{}_{|n} 	\,.
\end{equation}
Thus, we have
\begin{equation}  \label{div-a}
	\operatorname{div}\boldsymbol{\mathsf{a}}=\mathbf{0} 	\,.
\end{equation}
Using \eqref{div-a} in \eqref{Equilibrium-a}, one concludes that
\begin{equation}  \label{Geodesic-Equation}
	\nabla_{\boldsymbol{\mathsf{a}}}\,\boldsymbol{\mathsf{a}}=\mathbf{0} \,.
\end{equation}
This is the geodesic equation that tells us that the integral curves of $\boldsymbol{\mathsf{a}}$ are geodesics. But we know that in the Euclidean space geodesics are straight lines. This in turn implies that the stress is homogeneous.\\
Case \textit{(iii)}: The three eigenvalues are distinct. In this case, the stress has the following representation \eqref{Stress-Specrtal} and we have
\begin{equation} 
	\operatorname{div}\boldsymbol{\sigma}= 
	(\sigma_1-\sigma_3) \,\nabla_{\boldsymbol{\mathsf{a}}}\,\boldsymbol{\mathsf{a}}	
	+(\sigma_2-\sigma_3) \,\nabla_{\boldsymbol{\mathsf{b}}}\,\boldsymbol{\mathsf{b}}
	=\mathbf{0} \,.
\end{equation}
In components
\begin{equation} \label{Equilibrium-a-b}
	(\sigma_1-\sigma_3) \,\mathsf{a}^m{}_{|n}\mathsf{a}^n	
	+(\sigma_2-\sigma_3) \,\mathsf{b}^m{}_{|n}\mathsf{b}^n
	=0 \,.
\end{equation}
Hence, by taking the dot product of both sides of the above identity with $\boldsymbol{\mathsf{a}}$, one obtains
\begin{equation} 
	0=(\sigma_1-\sigma_3) \,\mathsf{a}_m\,\mathsf{a}^m{}_{|n}\mathsf{a}^n	
	+(\sigma_2-\sigma_3) \,\mathsf{a}_m\,\mathsf{b}^m{}_{|n}\mathsf{b}^n
	=(\sigma_2-\sigma_3) \,\mathsf{a}_m\,\mathsf{b}^m{}_{|n}\mathsf{b}^n=0 \,,
\end{equation}
and hence
\begin{equation}  \label{Identity-a-b}
	\mathsf{a}_m\,\mathsf{b}^m{}_{|n}\mathsf{b}^n=0\quad \text{or} \qquad
	\boldsymbol{\mathsf{a}}\cdot \nabla_{\boldsymbol{\mathsf{b}}}\,\boldsymbol{\mathsf{b}}=0
	 \,.
\end{equation}
Similarly
\begin{equation} \label{Identity-b-a}
	\boldsymbol{\mathsf{b}}\cdot \nabla_{\boldsymbol{\mathsf{a}}}\,\boldsymbol{\mathsf{a}}=0
	 \,.
\end{equation}
In summary, $\boldsymbol{\mathsf{a}}\cdot \nabla_{\boldsymbol{\mathsf{a}}}\,\boldsymbol{\mathsf{a}}=
	\boldsymbol{\mathsf{b}}\cdot \nabla_{\boldsymbol{\mathsf{a}}}\,\boldsymbol{\mathsf{a}}=
	\boldsymbol{\mathsf{a}}\cdot \nabla_{\boldsymbol{\mathsf{b}}}\,\boldsymbol{\mathsf{b}}=
	\boldsymbol{\mathsf{b}}\cdot \nabla_{\boldsymbol{\mathsf{b}}}\,\boldsymbol{\mathsf{b}}=
	0$.
Therefore
\begin{equation} 
	\nabla_{\boldsymbol{\mathsf{a}}}\,\boldsymbol{\mathsf{a}}=k_1\,\boldsymbol{\mathsf{c}}\,,\qquad
	\nabla_{\boldsymbol{\mathsf{b}}}\,\boldsymbol{\mathsf{b}}=k_2\,\boldsymbol{\mathsf{c}}
	 \,.
\end{equation}
Instead of the spectral representation \eqref{Stress-Specrtal} one can equivalently write
\begin{equation} \label{Stress-Specrtal1}
	\boldsymbol{\sigma}= 
	(\sigma_1-\sigma_2) \,\boldsymbol{\mathsf{a}}\otimes\boldsymbol{\mathsf{a}}	
	+(\sigma_3-\sigma_2) \,\boldsymbol{\mathsf{c}}\otimes\boldsymbol{\mathsf{c}}
	+\sigma_2 \,\mathbf{g}^\sharp	\,.
\end{equation}
Using this representation and following a similar argument that led to the identity \eqref{Identity-b-a}, one can show that $\boldsymbol{\mathsf{c}}\cdot \nabla_{\boldsymbol{\mathsf{a}}}\,\boldsymbol{\mathsf{a}}=0$, and hence $k_1=0$, i.e., $\nabla_{\boldsymbol{\mathsf{a}}}\,\boldsymbol{\mathsf{a}}=\mathbf{0}$.
One can also use the following spectral representation
\begin{equation} \label{Stress-Specrtal2}
	\boldsymbol{\sigma}= 
	(\sigma_2-\sigma_1) \,\boldsymbol{\mathsf{b}}\otimes\boldsymbol{\mathsf{b}}
	+(\sigma_3-\sigma_1) \,\boldsymbol{\mathsf{c}}\otimes\boldsymbol{\mathsf{c}}	
	+\sigma_1 \,\mathbf{g}^\sharp	\,.
\end{equation}
Using this representation and following a similar argument that led to the identity \eqref{Identity-a-b}, one can show that $\boldsymbol{\mathsf{c}}\cdot \nabla_{\boldsymbol{\mathsf{b}}}\,\boldsymbol{\mathsf{b}}=0$, and hence $k_2=0$, i.e., $\nabla_{\boldsymbol{\mathsf{b}}}\,\boldsymbol{\mathsf{b}}=\mathbf{0}$. 
From $\nabla_{\boldsymbol{\mathsf{a}}}\,\boldsymbol{\mathsf{a}}=\nabla_{\boldsymbol{\mathsf{b}}}\,\boldsymbol{\mathsf{b}}=\mathbf{0}$ we conclude that the integral curves of $\boldsymbol{\mathsf{a}}$ and $\boldsymbol{\mathsf{b}}$ are orthogonal straight lines. This implies that the integral curves of $\boldsymbol{\mathsf{c}}$ are straights lines as well and orthogonal to those of  $\boldsymbol{\mathsf{a}}$ and $\boldsymbol{\mathsf{b}}$. This again implies that the Cauchy stress is homogeneous. 
\end{proof}

\paragraph{Universal deformations for a special class of implicit-elastic solids.}
Next, we consider the special subclass of materials with constitutive equations \eqref{invertible-stress}, i.e., 
\begin{equation} 
	\mathbf{b}^\sharp=\xi_0\,\mathbf{g}^\sharp+\xi_1 \boldsymbol{\sigma}+\xi_2 \boldsymbol{\sigma}^2	\,.
\end{equation}
We take the divergence of both sides and use the fact that $\operatorname{div}\boldsymbol{\sigma}=\mathbf{0}$ (equilibrium equations in the absence of body forces) and  $\operatorname{div}\mathbf{g}^\sharp=\mathbf{0}$ (metric is covariantly constant):
\begin{equation} 
	\nabla\xi_0+ \boldsymbol{\sigma} \nabla\xi_1+ \boldsymbol{\sigma}^2 \nabla\xi_2
	+\xi_2 \operatorname{div}\boldsymbol{\sigma}^2 -\operatorname{div}\mathbf{b}^\sharp=0\,.
\end{equation}
The above identity holds for arbitrary response functions $\xi_0, \xi_1, \xi_2$. Let us replace $\xi_0$ by $\zeta \xi_0$: $\zeta\nabla\xi_0+ \boldsymbol{\sigma} \nabla\xi_1+ \boldsymbol{\sigma}^2 \nabla\xi_2	+\xi_2 \operatorname{div}\boldsymbol{\sigma}^2 -\operatorname{div}\mathbf{b}^\sharp=0$. Differentiating both sides of the above identity with respect to $\zeta$ one obtains $\nabla\xi_0=0$, and hence the principal invariants of Cauchy stress are constant. Thus, Cauchy stress is homogeneous and $\xi_0, \xi_1, \xi_2$ are constant. Hence, $\mathbf{b}^\sharp$ is homogeneous. Therefore, universal deformations are homogeneous.
In Example \ref{Ex:quadratic} we saw that not every homogeneous deformation is admissible for this class of elastic materials.
In other words, unlike Cauchy (and Green) elasticity, the set of universal deformations depends on the response functions. All we know is that for any material within this class the set of universal deformations is a subset of homogeneous deformations.

\begin{remark}
To prove that universal deformations in homogeneous compressible isotropic hyperelastic solids are homogeneous, \citet{Ericksen1955} used the equilibrium equations in the absence of body forces, the necessity for these equations to hold for any energy function, and the compatibility equations of the left Cauchy-Green strain. A simpler proof was presented in \citep{Saccomandi2001b} by employing the following two specific energy functions: $W(I_1,I_2,I_3)=\mu_1I_1+\mu_2I_2+\mu_3I_3$ and $\hat{W}(I_1,I_2,I_3)=\hat{\mu}_1I_1^2+\hat{\mu}_2I_2^2+\hat{\mu}_3I_3^2$, where $\mu_i$ and $\hat{\mu}_i$, $i=1,2,3$, are arbitrary constants. \citet{Casey2004} demonstrated that Ericksen's result can be proved using the simpler energy functions $W=\mu_1I_1$ and $\hat{W}=\hat{\mu}_1I_1^2$. 
Furthermore, he concluded that universal deformations for ``all homogeneous elastic materials" must be homogeneous. While this is true, it is important to note that not every homogeneous deformation is admissible for any given elastic material, as demonstrated in Example \ref{Ex:quadratic}.
\end{remark}

\paragraph{Universal deformations in implicit-elastic solids.}
Universal deformations correspond to those deformations that satisfy the equilibrium equations in the absence of body forces for arbitrary constitutive functions $\alpha_i,~i=0,\hdots,8$.
We pick an arbitrary deformation, which has a left Cauchy-Green strain $\mathbf{b}^\sharp$ (and its corresponding $\mathbf{c}^\sharp$). The implicit constitutive equation \eqref{Implicit-Constitutive-Equation-Isotropic} determines the corresponding stress $\boldsymbol{\sigma}$ (that may not be unique) if $\mathbf{b}^\sharp$ is admissibe. 
Note that for this fixed deformation, $\boldsymbol{\sigma}=\boldsymbol{\sigma}(\alpha_0,\hdots,\alpha_8)$.
This means that when we choose different materials within the class of materials \eqref{Implicit-Constitutive-Equation-Isotropic}, $\boldsymbol{\sigma}$ changes. In other words, a given deformation has, in general, infinitely many corresponding stresses within the material class. 
Any stress $\boldsymbol{\sigma}$ corresponding to the fixed $\mathbf{b}^\sharp$ and $\mathbf{c}^\sharp$ must satisfy \eqref{Implicit-Constitutive-Equation-Isotropic} everywhere in the body.
We take the (spatial) divergence of both sides of \eqref{Implicit-Constitutive-Equation-Isotropic} and note that $\operatorname{div}\boldsymbol{\sigma}=\mathbf{0}$ (equilibrium equations in the absence of body forces) and  $\operatorname{div}\mathbf{g}^\sharp=\mathbf{0}$ (metric is covariantly constant):
\begin{equation} \label{Equilibrium-ICE}
\begin{aligned}
	&\nabla\alpha_0
	+ \boldsymbol{\sigma}\cdot \nabla\alpha_1
	+ \boldsymbol{\sigma}^2\cdot \nabla\alpha_2
	+\alpha_2 \operatorname{div}\boldsymbol{\sigma}^2
	+\alpha_3\operatorname{div}\mathbf{b}^\sharp
	+\mathbf{b}^\sharp\cdot \nabla\alpha_3
	+\alpha_4\operatorname{div}\mathbf{c}^{\sharp} 
	+\mathbf{c}^\sharp\cdot \nabla\alpha_4 \\
	&+\alpha_5 \left[ \nabla\boldsymbol{\sigma}\cdot\mathbf{b}^\sharp
	+\nabla\mathbf{b}^\sharp \cdot \boldsymbol{\sigma} 
	+\boldsymbol{\sigma}\operatorname{div}\mathbf{b}^\sharp
	\right]
	+\left(\boldsymbol{\sigma}\mathbf{b}^\sharp
	+\mathbf{b}^\sharp\boldsymbol{\sigma}\right)\cdot \nabla\alpha_5\\
	&+\alpha_6 \left[ \nabla\boldsymbol{\sigma}^2\cdot\mathbf{b}^\sharp
	+\nabla\mathbf{b}^\sharp \cdot \boldsymbol{\sigma}^2 
	+\boldsymbol{\sigma}^2\operatorname{div}\mathbf{b}^\sharp
	+\mathbf{b}^\sharp\operatorname{div}\boldsymbol{\sigma}^2
	\right]
	+\left(\boldsymbol{\sigma}^2\mathbf{b}^\sharp
	+\mathbf{b}^\sharp\boldsymbol{\sigma}^2\right)\cdot \nabla\alpha_6\\
	&+\alpha_7 \left[ \nabla\boldsymbol{\sigma}\cdot\mathbf{c}^\sharp
	+\nabla\mathbf{c}^\sharp \cdot \boldsymbol{\sigma} 
	+\boldsymbol{\sigma}\operatorname{div}\mathbf{c}^\sharp
	\right]
	+\left(\boldsymbol{\sigma}\mathbf{c}^\sharp
	+\mathbf{c}^\sharp\boldsymbol{\sigma}\right)\cdot \nabla\alpha_7\\
	&+\alpha_8 \left[ \nabla\boldsymbol{\sigma}^2\cdot\mathbf{c}^\sharp
	+\nabla\mathbf{c}^\sharp \cdot \boldsymbol{\sigma}^2 
	+\boldsymbol{\sigma}^2\operatorname{div}\mathbf{c}^\sharp
	+\mathbf{c}^\sharp\operatorname{div}\boldsymbol{\sigma}^2
	\right]
	+\left(\boldsymbol{\sigma}^2\mathbf{c}^\sharp
	+\mathbf{c}^\sharp\boldsymbol{\sigma}^2\right)\cdot \nabla\alpha_8=\mathbf{0}\,.
\end{aligned}
\end{equation}
For fixed $\mathbf{b}^\sharp$ and $\mathbf{c}^\sharp$, $\boldsymbol{\sigma}=\boldsymbol{\sigma}(\alpha_0,\hdots,\alpha_8)$, and the above identity must hold everywhere in the deformed body for arbitrary response functions $\alpha_i=\alpha_i(I_1,\hdots,I_{10})$, $i=0,\hdots,8$. 
Note that $\boldsymbol{\sigma}$ and $\boldsymbol{\sigma}^2$ do not depend on the derivative of $\alpha_0$ with respect to any of the ten invariants. 
Thus
\begin{equation}
	\nabla\alpha_0=\sum_{j=1}^{10} \frac{\partial \alpha_0}{\partial I_j}\,\nabla I_j=\mathbf{0}\,.
\end{equation}
Therefore, the arbitrariness of the derivatives of $\alpha_0$ with respect to the ten invariants implies that 
\begin{equation} \label{Invarinats-Constant}
	I_1,\hdots,I_{10}~\text{~are~constant}\,.
\end{equation}
$I_1, I_2, I_3$ being constant tells us that the Cauchy stress is homogeneous. 
Thus, \eqref{Equilibrium-ICE} is simplified to read
\begin{equation} \label{Equilibrium-ICE1}
\begin{aligned}
	&\alpha_3\operatorname{div}\mathbf{b}^\sharp
	+\alpha_4\operatorname{div}\mathbf{c}^{\sharp} 
	+\alpha_5\, \boldsymbol{\sigma}\operatorname{div}\mathbf{b}^\sharp
	+\alpha_6 \left[ \nabla\mathbf{b}^\sharp \cdot \boldsymbol{\sigma}^2 
	+\boldsymbol{\sigma}^2\operatorname{div}\mathbf{b}^\sharp
	\right]\\
	&
	+\alpha_7 \left[ \nabla\mathbf{c}^\sharp \cdot \boldsymbol{\sigma} 
	+\boldsymbol{\sigma}\operatorname{div}\mathbf{c}^\sharp
	\right]
	+\alpha_8 \left[\nabla\mathbf{c}^\sharp \cdot \boldsymbol{\sigma}^2 
	+\boldsymbol{\sigma}^2\operatorname{div}\mathbf{c}^\sharp
	\right]
	=\mathbf{0}\,.
\end{aligned}
\end{equation}
In the constitutive equation \eqref{Implicit-Constitutive-Equation-Isotropic} let us assume that $\alpha_0$, $\alpha_1$, $\alpha_3$, and $\alpha_4$ are the only nonzero response functions (this corresponds to Cauchy elasticity). Thus, in the above identity the response function $\alpha_3$ and $\alpha_4$ can be chosen arbitrarily, and hence $\operatorname{div}\mathbf{b}^\sharp=\operatorname{div}\mathbf{c}^\sharp=\mathbf{0}$.
We observe that the universality constraints of implicit elasticity include those of Cauchy elasticity, i.e., \eqref{Universality-Constraints-Cauchy}. Therefore, universal deformations are homogeneous, and \eqref{Equilibrium-ICE1} is now trivially satisfied. 
Therefore we have proved the following result.

\begin{prop}
In compressible isotropic implicit elasticity, all universal deformations are homogeneous. 
\end{prop}

\begin{remark}
It should be emphasized that for a given implicit-elastic material not every homogeneous deformation is admissible. In other words, the set of universal deformations is material dependent, in general. All one knows is that for a given material within the class of implicit-elastic materials, the set of universal deformations is a subset of homogeneous deformations. 
This suggests that homogeneous deformations remain useful for characterizing material properties of implicit-elastic materials.
\end{remark}

\section{Conclusions}  \label{Sec:Conclusions}

Universal deformations have been extensively studied for compressible and incompressible hyperelastic solids. 
In recent years, the study of universal deformations has been expanded to encompass anelasticity, as well as anisotropic and inhomogeneous hyperelastic bodies. More recently, universal deformations have been studied for compressible and incompressible Cauchy elasticity, which includes hyperelasticity as a subset.
We also know that there are larger classes of elastic solids that include, for instance, Cauchy elasticity. One such class of solids was introduced by \citet{Morgan1966} characterized by implicit constitutive equations of the form $\boldsymbol{\mathsf{f}}(\boldsymbol{\sigma},\mathbf{b})=\mathbf{0}$. The study of this class of elastic solids has been revived in the past twenty years by Rajagopal and his collaborators \citep{Rajagopal2003,Rajagopal2007,Bustamante2009,Bustamante2011}.
In this paper we investigated the problem of finding the universal deformations of compressible isotropic solids with the implicit constitutive equations $\boldsymbol{\mathsf{f}}(\boldsymbol{\sigma},\mathbf{b})=\mathbf{0}$.
We showed that the universal deformations for these materials are homogeneous. 

However, we observed that, in general, not every homogeneous deformation is admissible for a given implicit-elastic material. This implies that in implicit elasticity the set of universal deformations is material dependent. However, for any given implicit-elastic material, the set of universal deformations is a subset of the set of homogeneous deformations. This suggests that homogenous deformations can still be used to characterize the material properties of such solids.

\section*{Acknowledgments}

This work was supported by NSF -- Grant No. CMMI 1939901.

\bibliographystyle{abbrvnat}
\bibliography{ref,ref1}

\end{document}